\documentclass[12pt, prd, showpacs]{revtex4}
\usepackage{amssymb}
\usepackage{amsmath}

\input{tcilatex}

\begin{document}

\title{Special case of the Ba\~{n}ados-Silk-West effect }
\author{O. B. Zaslavskii}
\affiliation{Department of Physics and Technology, Kharkov V.N. Karazin National
University, 4 Svoboda Square, Kharkov 61022, Ukraine}
\affiliation{Institute of Mathematics and Mechanics, Kazan Federal University, 18
Kremlyovskaya St., Kazan 420008, Russia}
\email{zaslav@ukr.net }

\begin{abstract}
If two particles collide near the rotating extremal black hole and one of
them is fine-tuned, the energy in the center of mass frame $E_{c.m.}$ can
grow unbounded. This is the so-called Ba\~{n}ados-Silk-West (BSW) effect.
Recently, another type of high energy collisions was considered in which all
processes happen in the Schwarzschild background with free falling
particles. If the Killing energy $E$ of one of particle is sufficiently
small, $E_{c.m.}$ grows unbounded. We show that, however, such a particle
cannot be created in any precedent collision with finite energies, angular
momenta and masses. Therefore, in contrast to the standard BSW effect, this
one cannot be realized if initial particles fall from infinity. If the black
hole is electrically charged, such a type of collisions is indeed possible,
when a particle with very small $E$ collides with one more particle coming
from infinity. Thus the BSW effect is achieved due to collisions of neutral
particles in the background of a charged black hole. This requires, however,
at least two-step process.
\end{abstract}

\maketitle

\section{Introduction}

If two particles collide near a rotating black hole, under certain condition
this leads to the unbounded growth of the energy in the center of mass frame 
$E_{c.m.}$ This is the essence of the so-called Ba\~{n}ados-Silk-West (BSW)
effect \cite{ban} (see also more early works \cite{pir1} - \cite{pir3}). The
aforementioned condition requires that one of colliding particles has
fine-tuned parameters, so that $X\equiv E-\omega L$ vanishes or is
sufficiently small near the horizon. Here, $E$ is the Killing energy, $L$
being the angular momentum, $\omega $ a certain metric coefficient
responsible for rotation. There is also a static charged counterpart of the
BSW effect \cite{jl}, when the aforementioned condition reads $E-q\varphi
\approx 0$ near the horizon. Here, $q$ is the particle's charge, $\varphi $
being the Coulomb potential of a black hole. When there is neither rotation
nor electric charge, the effect, in general, disappears. In particular, if
two particles of equal masses with $E_{1}=E_{2}=m$, collide in the
Schwarzschild background, $E_{c.m.}\leq 2\sqrt{5}m$ \cite{baus}. Hereafter,
we use subscript "i" to indicate quantities related to particle $i$.

Meanwhile, there is a rather special case when the criticality condition can
be formally satisfied even without rotation or the electric charge. This
happens if $E\approx 0$ itself is small that can be satisfied even for the
Schwarzschild black hole. This observation was made in \cite{gp11} (see
discussion on p. 3864 before eq. 35 there). Quite recently, it was
rediscovered in \cite{eva}. It is worth noting that the criticality
condition $E=0$ was also considered in Sec. II of \cite{ots} for 2+1 black
holes with the cosmological constant \cite{btz}. It appeared as a limit of
the angular momentum $L\rightarrow 0$ in a more general condition (2.15)
there. The corresponding space-time is not asymptotically flat but near the
horizon the same features manifest themselves, so $E_{c.m.}$ grows unbounded
when $E\rightarrow 0$ for one of two particles.

Meanwhile, there is a problem with physical realization of such a scenario.
The corresponding particle cannot come from infinity, where $E\geq m$, $m$
being the particle mass. So small energy can be obtained if a particle is
maintained near the horizon. Indeed, the energy of \ a particle that remains
in the rest in the static field $E=m\sqrt{-g_{00}}$, where $g_{00}$ is the
corresponding component of the metric tensor. If, in the Schwarzschild
metric, $g_{00}\rightarrow 0$, the energy $E\rightarrow 0$ as well. However,
if such a particle is kept fixed, the experienced acceleration $a\sim 1/l$,
where $l$ is the proper distance to the horizon. Letting such a particle
move freely, one indeed obtains unbounded $E_{c.m.}$ after its collision
with some other particle. But this result is gained by the expense of
unbounded forces that were exerted on a particle before it has been
released. To a large extent, this deprives the scenario under discussion of
physical significance.

If, nonetheless, we want some scenario with small $E$ to be more physical,
we must elucidate, whether or not it can be realized by means of particles
that move along geodesics or under the action of a finite force. This
problem is considered in the present work. It was conjectured in the end of
Ref. \cite{eva} that particle in question can be obtained in some foregoing
collision. We show that for the Schwarzschild black hole this is impossible.
Instead, this is indeed possible in the background of the extremal
Reissner-Nordstr\"{o}m (RN) black hole.

The paper is organized as follows. In Sec. \ref{eq} we give the form of the
metric and equations of motion in the RN metric. In Sec. \ref{col} we
consider collisions of particles 1 and 2 that turn into particles 3 and 4
and list basic formulas for the energy in the center of mass frame. In Sec. %
\ref{spec} we discuss the collision in the special case when particles are
neutral. In Sec. \ref{ov} we list basic formulas that enable us to find
dynamic characteristics of new particles, given the data of initial ones.
These formulas are exact. In Sec. \ref{neut} we elucidate, whether we can
obtain the energy $E_{3}$ of particle 3 as small as we like. The answer is
negative. Further, we discuss the second collision of neutral particles when
particle 5 comes form infinity. In Sec. \ref{extremal} we consider processes
in the extremal RN background. The combined fine-tuned (critical) particle 0
(equivalent to 1+2) decays to particles 3 and 4. We show that we can obtain
a neutral particle 3 with very small $E_{3}$ that leads to the analogue of
the BSW effect in next collision with particle 5. In Sec. \ref{concl} we
summarize the results.

In what follows, we use the geometric system of units in which fundamental
constants $G=c=1$.

\section{Equations of motion\label{eq}}

Let us consider the black hole metric%
\begin{equation}
ds^{2}=-dt^{2}f+\frac{dr^{2}}{f}+r^{2}d\omega ^{2}\,\text{,}  \label{sph}
\end{equation}%
where $d\omega ^{2}=d\theta ^{2}+\sin ^{2}\theta d\phi ^{2}$, $f=f(r).$ The
largest root $r=r_{+}$ of equation $f=0$ corresponds to the event horizon.
If a particle moves along the electrogeodesic (i.e. under the action of
gravitation and electrostatic force only), for motion within the equatorial
plane we have%
\begin{equation}
m\dot{t}=\frac{X}{f}\text{,}  \label{t}
\end{equation}%
\begin{equation}
m\dot{\phi}=\frac{L}{r^{2}}\text{,}
\end{equation}%
\begin{equation}
X=E-q\varphi \text{,}  \label{x}
\end{equation}%
\begin{equation}
m\dot{r}=\sigma P\text{, }P=\sqrt{X^{2}-f\tilde{m}^{2}}\text{, }\tilde{m}%
^{2}=m^{2}+\frac{L^{2}}{r^{2}},  \label{r}
\end{equation}%
dot denotes derivative with respect to the proper time, $\sigma =\pm 1$. The
forward-in-time condition $\dot{t}>0$ entails%
\begin{equation}
X\geq 0\text{.}  \label{ftw}
\end{equation}

In the case of the Schwarzschild black hole, $\varphi =0$, $f=1-\frac{r_{+}}{%
r}$. For the extremal RN black hole, $\varphi =\frac{r_{+}}{r}$, $f=\left( 1-%
\frac{r_{+}}{r}\right) ^{2}$, so 
\begin{equation}
X=E-q+q\sqrt{f}\text{.}  \label{ext}
\end{equation}

In what follows, we use the standard terminology. If $X_{H}=0$, a particle
is \ called critical. This is realized if $E=q$. If $X_{H}$ is separated
from zero, it is called usual. If $X_{H}=O(\sqrt{f}_{c})$, it is called
near-critical. Here, subscript "H" refers to the quantity calculated on the
horizon and "c" to that taken in the point of collision.

\section{Particle collisions and energy in the center of mass frame \label%
{col}}

If particles 1 and 2 collide, for the energy in the center of mass frame we
have%
\begin{equation}
E_{c.m.}^{2}=-(m_{1}u_{1\mu }+m_{2}u_{2\mu })(m_{1}u_{1}^{\mu
}+m_{2}u_{2}^{\mu })=m_{1}^{2}+m_{2}^{2}+2m_{1}m_{2}\gamma \text{,}
\label{cm}
\end{equation}%
where $\gamma =-u_{1\mu }u_{2}^{\mu }$ is the relative Lorentz factor of
particle motion, $u_{1,2}^{\mu }$ are the four-velocities of particles.
Then, for motion within the equatorial plane one obtains from (\ref{t}) - (%
\ref{r}) that for motion of both particles in the same direction%
\begin{equation}
m_{1}m_{2}\gamma =\frac{X_{1}X_{2}-P_{1}P_{2}}{f}-\frac{L_{1}L_{2}}{r^{2}}%
\text{.}  \label{ga}
\end{equation}

If collision happens near the horizon, particle 1 is critical and particle 2
is usual, we have unbounded growth of $\gamma $ \cite{jl}%
\begin{equation}
E_{c.m.}^{2}\approx 2\frac{\left( X_{2}\right) _{H}(E_{1}-\sqrt{E_{1}^{2}-%
\tilde{m}_{1}^{2}})}{\sqrt{f}}\text{.}
\end{equation}

\section{Special scenario \label{spec}}

Let us consider collision of neutral particles, so all $q_{i}=0,$ $%
X_{i}=E_{i}$. Then, in (\ref{ga}), $P_{i}=\sqrt{E_{i}^{2}-f\tilde{m}_{i}^{2}}
$. We also assume that $E_{1}$ is very small,%
\begin{equation}
E_{1}=\alpha \sqrt{f_{c}}\text{,}
\end{equation}%
where $\alpha $ is some constant, $f_{c}\ll 1$ for collision near the
horizon. Then,%
\begin{equation}
m_{1}m_{2}\gamma \approx \frac{E_{2}(\alpha -\sqrt{\alpha ^{2}-\tilde{m}%
_{1}^{2}})}{\sqrt{f_{c}}}
\end{equation}%
formally grows unbounded. This just corresponds to the situation described
in \cite{gp11} - \cite{ots}. However, we want particle 1 to be created in
some precedent collision, preferably due to the process that involves
particles coming from infinity. As, in the case under consideration, they
both are usual, the first collision must occur with finite $E_{c.m.}$

As is already established \cite{k}, the BSW effect has a simple explanation.
The quantity $X$ (\ref{x}) obeys the relation%
\begin{equation}
X=m\frac{\sqrt{f}}{\sqrt{1-V^{2}}}\text{,}  \label{xn}
\end{equation}%
where $V$ is the three-velocity measured by a static observer (see eq. 29 in
the aforementioned work). Then, for a usual particle, the horizon limit $%
f\rightarrow 0$ shows that $V\rightarrow 1$. Meanwhile, for the critical or
near-critical particle, the left hand side of (\ref{xn}) has the order $%
\sqrt{f}$, so this equation is satisfied with $V<1$. Then, collision of a
rapid and slow particles results in the large relative velocity close to the
speed of light, and $\gamma $ grows unbounded.

The case under consideration has its specific feature. Now, $q=0$, $X=E$.
Therefore, the left hand side can be made as small as one likes not due to
fine tuning between parameters $E$ and $q$ but due to small value of $E=O(%
\sqrt{f})$ itself.

\section{The first collision and the overall scheme \label{ov}}

We assume that after collision, new particles 3 and 4 appear. Alternatively,
we can consider decay of particle 0 that formally combines particle 1 and 2.
The conservation laws in the point of collision tell us%
\begin{equation}
E_{0}=E_{1}+E_{2}=E_{3}+E_{4}\text{,}
\end{equation}%
\begin{equation}
L_{0}=L_{1}+L_{2}\text{,}
\end{equation}%
\begin{equation}
q_{0}=q_{1}+q_{2}=q_{3}+q_{4}\text{,}
\end{equation}%
\begin{equation}
-P_{0}=-P_{1}-P_{2}=\sigma _{3}P_{3}+\sigma _{4}P_{4}\text{.}
\end{equation}%
It follows from these equations that%
\begin{equation}
X_{0}=X_{1}+X_{2}=X_{3}+X_{4}\text{.}
\end{equation}

Now, it is convenient to take advantages of the results already obtained in
the previous work \cite{centr} and listed there in eqs. (19) - (25). The
only obvious difference is that now instead of $\omega L$, the quantity $X$
contains $q\varphi $ (\ref{x}). If particle 0 is thought of as a combined
one, $m_{0}$ coincides with the $E_{c.m.}$ in the particle collision.
Otherwise, $m_{0}$ is simply the mass of particle $0$. Then, straightforward
algebraic manipulation give us

\begin{equation}
\left( X_{3}\right) _{c}=\frac{1}{2\tilde{m}_{0}^{2}}(X_{0}\Delta _{+}+P_{0}%
\sqrt{D}\delta )_{c}\text{,}  \label{3}
\end{equation}%
\begin{equation}
\left( X_{4}\right) _{c}=\frac{1}{2\tilde{m}_{0}^{2}}(X_{0}\Delta _{-}-P_{0}%
\sqrt{D}\delta )_{c}\text{,}
\end{equation}%
where $\delta =1$ or $\delta =-1$. 
\begin{equation}
\Delta _{\pm }=\tilde{m}_{0}^{2}\pm (\tilde{m}_{3}^{2}-\tilde{m}_{4}^{2}).
\end{equation}%
The positivity of $X_{3.4}$ entails 
\begin{equation}
\Delta _{\pm }>0.
\end{equation}%
\begin{equation}
D=\Delta _{+}^{2}-4\tilde{m}_{0}^{2}\tilde{m}_{3}^{2}=\Delta _{-}^{2}-4%
\tilde{m}_{0}^{2}\tilde{m}_{4}^{2}.  \label{D}
\end{equation}%
It is necessary that%
\begin{equation}
D\geq 0,
\end{equation}%
\begin{equation}
\tilde{m}_{0}\geq \tilde{m}_{3}+\tilde{m}_{4}\text{.}  \label{m34}
\end{equation}

For charged particles, we have 4 conservation laws for 6 unknowns $E_{3,4}$, 
$L_{3,4}$, $q_{3,4}$. In the above formulas, all quantities related to
particles 1 and 2 (hence, those of effective particle 0 as well) are fixed.
We also assume that masses $m_{3,4}$ are fixed for any given process.
Meanwhile, one of two angular momentum (say, $L_{3}$) and one of charges
(say, $q_{3}$) remain free parameters.

Below, we are interested in the two-step scenario that, overall, can be
described as follows.

1) The first step. Two particles come from infinity, collide and produce two
new ones. The energy $E_{c.m.}$ is finite in the point of collision.

2) One of new particles (say, particle 3) has a very small energy $E_{3\text{%
.}}$

3) In point 2 that is more close to the horizon ($f_{2}<f_{1}$) it collides
with one more particle 5 coming from infinity. $E_{c.m.}$ in the second
event (collision between particles 3 and 5) is unbounded.

In this scheme, the first step can be replaced with the decay of one
particle 0 instead of collisions of two ones.

\section{Collisions of neutral particles \label{neut}}

\subsection{Generic subcase}

Let all particles be electrically neutral, so all $q_{i}=0,$ $X_{i}=E_{i}$.
If particle 0 comes from infinity, it is usual, since $E_{0}\geq m_{0}$. If
it is a combined particle, this is true as well, since $E_{0}\geq m_{1}+m_{2}
$. Now we ask, is it possible to achieve indefinitely small $X_{3,4}=E_{3,4}$%
, assuming that $m_{0}$ is finite and nonzero? If yes, this would mean that
decay of particle 0 to 3 and 4 leads to particle (say, 3) with indefinitely
small $E_{3}$. Then, the second collision between particle 3 and some
additional particle 5 coming from infinity would give us indefinitely large $%
E_{c.m.}$ as is explained above and was considered in \cite{gp11} - \cite%
{ots}.

For the candidate particle 3, we take $\delta =-1$ in (\ref{3}), since we
want to make $E_{3}$ (almost) zero. Using (\ref{m34}), it is \ convenient to
rewrite (\ref{3}) 
\begin{equation}
E_{3}=\frac{2\tilde{m}_{3}^{2}P_{0}^{2}+\frac{f}{2}\Delta _{+}^{2}}{%
(E_{0}\Delta _{+}+P_{0}\sqrt{D})},  \label{e3}
\end{equation}%
where we put $X_{0}=E_{0}$.

As particle $0$ is usual, it is seen from (\ref{r}) that in the horizon
limit $f\rightarrow 0$, the quantity $P_{0}\rightarrow X_{0}$. Thus the
numerator tends to $2\left( \tilde{m}_{3}^{2}\right) _{c}E_{0}^{2}$ and does
not vanish for any $\tilde{m}_{3}\neq 0$. Then, we cannot achieve
indefinitely small $E_{3}$, so the scenario under discussion does not work.

\subsection{Special subcases}

Is it possible to achieve $E_{3}\rightarrow 0$ by taking a very small $%
\tilde{m}_{3}$ ? Let, at first, $\tilde{m}_{3}=0\,\ $exactly. From the
definition of $\tilde{m}$ (\ref{r}), it follows that $m_{3}=0$ and $L_{3}=0$%
. Then, $P_{3}=E_{3}$. If particle 3 collides one more time in point 2 with
some usual particle 5 coming from infinity, we have from (\ref{cm}), (\ref%
{ga}) 
\begin{equation}
\left( E_{c.m.}^{2}\right) _{2}=2E_{3}\frac{(E_{5}-\sqrt{E_{5}^{2}-\tilde{m}%
_{5}^{2}f_{2}})}{f_{2}}+m_{5}^{2}\text{,}  \label{s1}
\end{equation}%
where according to (\ref{e3}) $E_{3}\sim f_{1}$ is small.

In the horizon limit, we obtain that $E_{c.m.}^{2}$ is finite,%
\begin{equation}
\left( E_{c.m.}^{2}\right) _{2}\approx \frac{E_{3}\left( \tilde{m}%
_{5}^{2}\right) _{H}}{E_{5}}+m_{5}^{2}\text{,}
\end{equation}%
so again there is no BSW effect.

Instead, we may try to choose $\tilde{m}_{3}$ to be small but nonzero. In
turn, two different subcases should be considered separately. In doing so, $%
\tilde{m}_{3}^{2}\approx m_{3}^{2}+\frac{L_{3}}{r_{+}^{2}}=const.$

\subsubsection{Subcase a}

\begin{equation}
\tilde{m}_{3}^{2}f_{1}\ll E_{3}^{2}  \label{a}
\end{equation}

Then, we obtain from the general expressions (\ref{cm}), (\ref{ga}) that%
\begin{equation}
\left( E_{c.m.}^{2}\right) _{2}\approx \frac{\tilde{m}_{3}^{2}}{E_{3}}%
E_{5}+m_{5}^{2}\text{.}  \label{E2}
\end{equation}

Meanwhile, it follows from (\ref{e3}) that%
\begin{equation}
\frac{\tilde{m}_{3}^{2}}{E_{3}}\leq \frac{2\tilde{m}_{3}^{2}E_{0}\Delta _{+}%
}{2\tilde{m}_{3}^{2}E_{0}^{2}+\frac{f}{2}\Delta _{+}^{2}}\leq \frac{\Delta
_{+}}{E_{0}}\text{,}
\end{equation}%
where we put in the horizon limit $P_{0}\approx E_{0}$ and took into account
that $D\approx \Delta _{+}^{2}$ because of small $\tilde{m}_{3}^{2}$. Then,
it follows that (\ref{E2}) is finite, there is no BSW effect.

\subsubsection{Subcase b}

\begin{subequations}
\begin{equation}
\tilde{m}_{3}^{2}f_{1}=\beta ^{2}E_{3}^{2}.  \label{b}
\end{equation}

Here, $\beta $ is some coefficient $O(1)$. Then, it follows from Eqs. (\ref%
{cm}) and (\ref{ga}) that 
\end{subequations}
\begin{equation}
\left( E_{c.m.}^{2}\right) _{2}\approx \frac{2E_{5}}{f_{2}}(E_{3}-\sqrt{%
E_{3}^{2}-\tilde{m}_{3}^{2}f_{2}})+m_{5}^{2}\text{.}
\end{equation}

This can be rewritten as%
\begin{equation}
\left( E_{c.m.}^{2}\right) _{2}\approx \frac{E_{5}E_{3}F(y,\beta ^{2})}{f_{1}%
}+m_{5}^{2}  \label{e2b}
\end{equation}%
\begin{equation}
F(y,\beta ^{2})\equiv \frac{(1-\sqrt{1-\beta ^{2}y})}{y}=\frac{\beta ^{2}}{1+%
\sqrt{1-\beta ^{2}y}},  \label{F}
\end{equation}%
where $y=\frac{f_{2}}{f_{1}}$, $0\leq y$ $\leq 1$.

Obviously, the function $F$ is finite. It is monotonically increasing with $y
$ and changes from 
\begin{equation}
F(0,\beta ^{2})=\frac{\beta ^{2}}{2}  \label{F0}
\end{equation}%
on the horizon to%
\begin{equation}
F(1,\beta ^{2})=1-\sqrt{1-\beta ^{2}}\text{,}  \label{F1}
\end{equation}%
if the 2nd collision occurs practically in the same point immediately after
the 1st one.

It is seen from (\ref{e3}) that for our limit $f_{1}\rightarrow 0$,
condition (\ref{b}) is compatible with (\ref{e3}) for%
\begin{equation}
\tilde{m}_{3}^{2}\approx A^{2}f_{1}
\end{equation}%
only, where $A$ is a constant. Then, $E_{3}\approx \frac{A}{\beta }f_{1}$,
where we infer from (\ref{e3}) that%
\begin{equation}
A=\beta \frac{2E_{0}^{2}A^{2}+\frac{\left( \Delta _{+}^{2}\right) _{H}}{2}}{%
2E_{0}\left( \Delta _{+}\right) _{H}}.
\end{equation}

This quadratic equation can be solved easily,%
\begin{equation}
A=\frac{\left( \Delta _{+}\right) _{H}}{2E_{0}\beta }(1\pm \sqrt{1-\beta ^{2}%
})\text{.}
\end{equation}

In (\ref{e2b}), the numerator and denominator have the same order $f_{1}$,
so we obtain%
\begin{equation}
\left( E_{c.m.}^{2}\right) _{2}\approx \frac{A}{\beta }E_{5}F+m_{5}^{2}\text{%
.}
\end{equation}%
Again, it is finite, there is no BSW effect.

Thus having enumerated all possible subcases, we come to the conclusion
that, starting from particles with finite parameters, one cannot create a
suitable particle 3 to arrange the second collision with unbounded $\left(
E_{c.m.}^{2}\right) _{2}$.

In \cite{eva} (see the paragraph before eq.10 there), it was assumed by hand
that $E_{3}\sim \sqrt{f_{1}}$. However, the requirement of the finiteness of 
$E_{c.m.}=m_{0}$ in the precedent collision in which particle $E_{3}$ was
created, imposes severe restrictions. As we saw, they give rise to another
dependence $E_{3}\sim f_{1}$ that makes $E_{c.m.}$ finite in the second
collision.

\section{Process with charged critical particles \label{extremal}}

Now, we consider particle motion in the extremal RN background. Then, for
critical particle 0 we have according to (\ref{ext}),

\begin{equation}
X_{0}=E_{0}\sqrt{f},P_{0}=\sqrt{(E_{0}^{2}-\tilde{m}_{0}^{2})f}\text{.}
\label{crit}
\end{equation}

We assume that particle 0 (that effectively can model the combination of
particles 1+2) turns into particles 3 and 4. As the initial particle 0 is
critical, it follows from (\ref{ftw}) that near the horizon $X_{3}$ and $%
X_{4}$ are both small. More precisely, if they are created in the collision
with finite $E_{c.m.}=m_{0}$, they have the same order $\sqrt{f_{1}}$. We
want to elucidate, whether or not the BSW effect is possible in the second
collision, if particle 3 is uncharged, $q_{3}=0$.

It follows from (\ref{e3}) that contains now $X_{0}$ in the denominator
instead of $E_{0}$ and (\ref{crit}) that%
\begin{equation}
E_{3}=C_{1}\sqrt{f_{1}}\text{,}
\end{equation}%
where $C_{1}=C(r_{1})$,%
\begin{equation}
C=\frac{2\tilde{m}_{3}^{2}(E_{0}^{2}-\tilde{m}_{0}^{2})+\frac{\Delta _{+}^{2}%
}{2}}{E_{0}\Delta _{+}+\sqrt{E_{0}^{2}-\tilde{m}_{0}^{2}}\sqrt{D}}\text{,}
\end{equation}%
\begin{equation}
P_{3}=\sqrt{f_{1}}\sqrt{C^{2}-\tilde{m}_{3}^{2}\frac{f}{f_{1}}}\text{.}
\end{equation}

For the process near the horizon, $C_{1}\approx C_{H}.$ If particle 3
collides with some usual particle 5,%
\begin{equation}
\left( E_{c.m.}^{2}\right) _{2}\approx \frac{2\left( X_{5}\right)
_{H}C_{H}F(y,\delta )}{\sqrt{f_{1}}}\text{,}
\end{equation}%
where again $y=\frac{f_{2}}{f_{1}}$ but now now $\delta =\frac{\tilde{m}%
_{3}^{2}}{C_{H}^{2}}$. The function $F$ is defined above in (\ref{F}), so (%
\ref{F0}) and (\ref{F1}) are still valid, the function $F$ is bounded.
However, because of small $f_{1}$ in the denominator, $\left(
E_{c.m.}^{2}\right) _{2}$ can be made unbounded.

The dependence $\left( E_{c.m.}^{2}\right) $ $\sim $ $f^{-1/2}$ is exactly
the same as in the case of the standard BSW effect for charged black holes 
\cite{jl}. However, there is a qualitative difference now. In the standard
case, one of two colliding particles should be electrically charged. This is
necessary to satisfy the criticality condition $X_{H}=0$. Meanwhile, now a
particle that plays in the second collision the same role as the
(near)critical particle does in the standard BSW effect, is \textit{neutral}%
. We see that the role of the electric charge in the problem is twofold. On
one hand, the charge is required for a black hole. More precisely, the
presence of the charge enables one to produce in the first event a particle
with $q_{3}=0$ and small $E_{3}$, as a result of collision between two
critical particles. In accordance with general rules, such a collision leads
to finite $E_{c.m.}$ (see kinematic explanation in Sec. III B of \cite{k}).
From the other hand, the charge is irrelevant on the second stage of the
process, when a neutral particle collides with a usual one coming from
infinity. Then, the BSW effect can be achieved even due to collision of two
neutral particles.

\section{Conclusions \label{concl}}

Thus we considered two complementary cases. 1) Particle motion occurs in the
neutral background. In particular, this is valid for the Schwarzschild
metric. This holds also for the RN one, provided all particles are
electrically neutral. Then, an initial particle 0 (true or effective
combined one) is usual. It is shown that, arranging the collision between
two initial particles coming from infinity, it is impossible to obtain
particle 3 with almost vanishing $E_{3}$ to realize the BSW effect in the
second collision. This means that the scenario in which particle 3 with $%
E_{3}\approx 0$ is created in the foregoing collision (as outlined in the
end of \cite{eva}) does not work for the Schwarzschild black hole. 2)
Particle 0 is critical. It decays to two fragments, one of which is neutral.
Then, the BSW effect is indeed possible in the next collision. This gives a
new type of the BSW effect for electrically charged black holes, realized
with the help of electrogeodesics and geodesics. The counterpart of the
phenomenon discussed in our paper should exist also for rotating neutral
black holes, then the role of particle 3 with aforementioned properties will
be played by a particle with $L_{3}=0$. For 2+1 dimensional black holes with
the cosmological constant it was realized in \cite{ots}. Meanwhile, one can
expect a similar phenomenon for 3+1 asymptotically flat black holes as well,
including the Kerr metric.

\begin{acknowledgments}
The work is performed according to the Russian Government Program of
Competitive Growth of Kazan Federal University.
\end{acknowledgments}

\end{document}